\documentclass{eas}
\usepackage{graphicx}
\begin{document}

\TitreGlobal{Mass Profiles and Shapes of Cosmological Structures}

\title{Intra-group Light in Hickson Compact Groups}
\author{Da Rocha, C.$^{1,}$}\address{Institut f\"ur Astrophysik G\"ottingen - 
Georg August Universit\"at G\"ottingen - Germany}
\address{Divis\~ao de Astrof\'{\i}sica, Instituto Nacional de Pesquisas 
Espaciais - DAS/INPE - Brazil}
\author{Mendes de Oliveira, C.}\address{Instituto de Astronomia, 
Geof\'{\i}sica e Ci\^encias Atmosf\'ericas - USP - Brazil}
\author{Ziegler, B. L.$^1$}
\runningtitle{Intra-group Light in HCGs}
\setcounter{page}{1}
\index{Da Rocha, C.}
\index{Mendes de Oliveira, C.}
\index{Ziegler, B. L.}

%
\begin{abstract}
We have analyzed the intra-group light component of 3 Hickson Compact
Groups (HCG 79, HCG 88 and HCG 95) with detections in two of them:
HCG 79, with $46\pm11\%$ of the total $B$ band luminosity and HCG 95
with $11\pm26\%$. HCG 88 had no component detected.  This component
is presumably due to tidally stripped stellar material trapped in
the group potential and represents an efficient tool to determine the
stage of dynamical evolution and to map its gravitational potential.
To detect this low surface brightness structure we have applied the
wavelet technique OV\_WAV, which separates the different components of
the image according to their spatial characteristic sizes.
\end{abstract}
\maketitle
\section{Introduction}
The diffuse intra-group light (IGL) component is a useful tool to measure
the intensity of the tidal interactions suffered by the galaxies and to
map the extension and shape of the groups' gravitational potential and
the dark matter halo.

In order to isolate the IGL we used the ``\`a trous'' wavelet transform
with a Multi-Scale Vision Model
(OV\_WAV -- Epit\'acio Pereira, Raba\c ca \& Da Rocha 2005; Da Rocha
\& Mendes de Oliveira 2005), which does not depend on ``a priori''
information.  The process detects different characteristic size
structures, separating the types of light source in the image.

\section{Observational Data and Analysis}
We have studied 3 Hickson compact groups (Hickson 1982), in different
evolutionary stages, (HCG 79, HCG 88 and HCG 95) as a pilot study for
an IGL survey.

Simulated images were analyzed with OV\_WAV and showed that we are able
to detect low-surface brightness extended structures, down to a $S/N =
0.1$ per pixel, which corresponds to a 5-$\sigma$-detection level in
wavelet space.

\section{Results and Conclusions}
We have detected IGL in HCG 79 and HCG 95.  HCG 79 has an irregular IGL
distribution, which closely matches the X-Ray distribution (Pildis et
al. 1995) and is bluer than the galaxies ((B-R) = 1.5), possibly a mix
of stripped material from the outer parts of the galaxies and blue dwarf
galaxies destruction.  HCG 95 has an almost spherical IGL distribution,
with colors typical of old stellar populations. Non-detection of IGL in
HCG 88 indicating an early stage of dynamical evolution.

We suggest an evolutionary sequence: HCG 79, in an advanced stage of
dynamical evolution; HCG 95, intermediate stage; and HCG 88, initial
epoch still without IGL.
The presence of an IGL component indicates gravitationally bound
configurations in which tidal encounters already stripped a considerable
fraction of mass from the member galaxies and an advanced stage of
dynamical evolution, providing a test for formation and evolution models
of groups.

We are conducting an IGL survey in HCGs. Results will be compared
with N-body simulations in order to assess the dynamical age of those
dense structures.

\begin{table}

\centering
\caption{Properties of the IGL component detected in our sample.
\label{tabres}}
\begin{tabular}{cccccc}

\hline
Group & \multicolumn{2}{c}{\% ($B$ and $R$)} & $<\mu>_B$ & $B_{IGL}$ & $(B-R)_0$ \\
\hline

HCG 79  & $46\pm11$\% & $33\pm11$\% & $24.8\pm0.16$ & $14.0\pm0.16$ & $0.86\pm0.22$ \\
HCG 95  & $11\pm26$\% & $12\pm10$\% & $27.3\pm0.30$ & $16.9\pm0.30$ & $1.75\pm0.34$ \\
\hline
\end{tabular}

\end{table}

%
%
\begin{figure}
\centering
\includegraphics[width=3.6cm]{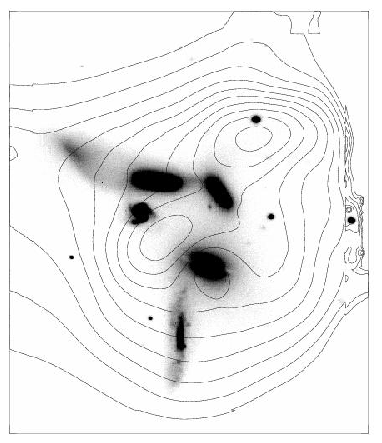}
\includegraphics[width=3.9cm]{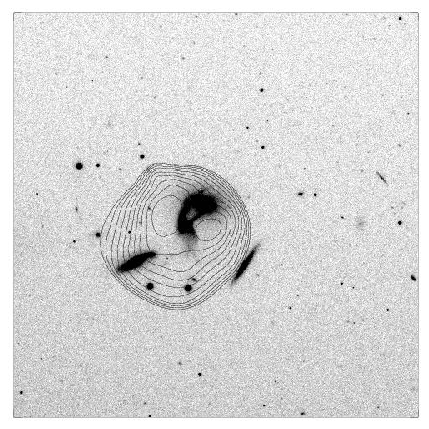}
\caption{B band image with IGL in contour curves superposed, ranging from
$24.2$ to $25.1$ magnitudes for HCG 79 (left) and from $26.9$ to $27.8$
magnitudes for HCG 95 (right).
\label{figdif}}
\end{figure}

\acknowledgements{
This project is supported by FAPESP (02/06881-4), CAPES/DAAD (BEX: 1380/04-4)
and VW Junior Research Group - Kinematic Evolution of Galaxies -
Volkswagen Foundation (I/76 520)}

\end{document}